\documentclass[
]{aa}
\usepackage{graphicx}
\usepackage{natbib}
\usepackage{txfonts}
\bibpunct{(}{)}{;}{a}{}{,}
\begin{document}
\title{Oxygen- and carbon-rich variable red giant populations in the 
Magellanic Clouds from\\ EROS, OGLE, MACHO, and 2MASS photometry.}
\titlerunning{Populations of variable red giants in the Magellanic Clouds}
\author{
M.~Wi\'{s}niewski\inst{1},
J.B.~Marquette\inst{2,3},
J.P.~Beaulieu\inst{2,3},
A.~Schwarzenberg-Czerny\inst{1,4},
P.~Tisserand\inst{5,6},
\'{E}.~Lesquoy\inst{6,2,3}
}
\authorrunning{Mariusz Wisniewski et al.}
\offprints{A. Schwarzenberg-Czerny}
\institute{Nicolaus Copernicus Astronomical Centre, Bartycka 18, 00-716 Warsaw, Poland
\and UPMC Universit\'e Paris 06, UMR7095, Institut d'Astrophysique de Paris, F-75014,
Paris, France, \email{(marquett,beaulieu,lesquoy)@iap.fr}
\and CNRS, UMR7095, Institut d'Astrophysique de Paris, F-75014, Paris, France
\and Adam Mickiewicz University Observatory, ul. S{\l}oneczna 36, PL 60-286, Pozna\'{n}, Poland, \email{alex@camk.edu.pl}
\and Research School of Astronomy \& Astrophysics, Mount Stromlo Observatory, Cotter Road, Weston ACT 2611, Australia, \email{tisseran@mso.anu.edu.au}
\and CEA, DSM, DAPNIA, Centre d'\'{E}tudes de Saclay, 91191 Gif-sur-Yvette Cedex, France
}
\date{Received \dots ; accepted \dots}
\abstract{The carbon-to-oxygen (C/O) ratio of asymptotic giant branch (AGB) stars
constitutes an important index of evolutionary and
environment/metallicity factor.}{We develop a method for
mass C/O classification of AGBs in photometric surveys without
using periods.}{For this purpose we rely on the slopes
in the tracks of individual stars in the colour-magnitude
diagram.}{We demonstrate that our method enables
the separation of C-rich and O-rich AGB stars with little confusion.
For the Magellanic Clouds we demonstrate that this method works for several
photometric surveys and filter combinations. As we rely on no
period identification, our results are relatively insensitive to the
phase coverage, aliasing, and time-sampling problems that plague
period analyses. For a subsample of our stars, we verify our C/O 
classification against published C/O catalogues. With our method we are able to
produce C/O maps of the entire Magellanic Clouds.}{Our purely
photometric method for classification of C- and O-rich AGBs
constitutes a method of choice for large, near-infrared
photometric surveys. Because our method depends on the slope of
colour-magnitude variation but not on magnitude zero point, it
remains applicable to objects with unknown distances.}
\keywords{Methods: data analysis, Techniques: photometric, Surveys,
 Stars: oscillations (including pulsations), Stars: AGB and post-AGB, Magellanic Clouds}
\maketitle

\section{Introduction}\label{s0}
Low and intermediate mass stars evolve through three late stages.
After passing the red giant branch (RGB) they reach maximum
luminosity at its tip (TRGB) followed by a luminosity drop after the
helium flash. Next they grow again and move along the asymptotic
giant branch (AGB). According to their spectra, the AGB stars
split into oxygen-rich stars (O-rich, M-stars, or K-stars) and
carbon-rich stars (C-rich, C-stars, or N-stars). The M-stars have
more oxygen than carbon in their atmospheres (C/O $<$ 1). When the abundance of oxygen
equals that of carbon the AGB type is S. Classification of the latter
is difficult and also involves intermediate MS and SC types
\citep{2003A&A...402..133C}.

Stars entering the AGB phase are rich in oxygen, however,
subsequent dredge-up caused by thermal pulsation may enrich their
atmospheres in carbon \citep{1983ARA&A..21..271I}. On the AGB, several
thermal pulses (TP-AGB) may occur, resulting in dredging up the
matter enriched in carbon nuclei by nuclear fusion. In this way
the O-rich stars are converted after at least several pulses 
into the C-rich ones (\citealt{2008A&A...482..883M} and references there).

It is believed that the rate of conversion of O-stars into C-stars
depends on the efficiency of the third dredge-up and the extent
and time variation in the massloss (e.g. \citealt{1981ApJ...246..278I}; \citealt{1999A&A...344..123M}). 
Mass loss is expected to be stronger in metal-rich
stars, giving to shorter AGB life and eventually yielding a C-star. These
thermal pulses result in the luminosity variations with the
peak-to-peak amplitude up to a few magnitudes at visual
wavelengths.

According to \citet{1983ARA&A..21..271I}, the lower the metallicity the less
carbon needed to be dredged-up in order to convert an O-star
into a C-star, hence the correlation between metallicity and the
C/M ratio. For lower metallicity, the AGB evolutionary tracks move
to higher temperatures; for very low metallicity a
post-horizontal-branch star may become a white dwarf without
first becoming an AGB star. Since O and C stars reveal very
different spectra, it is relatively easy to spectroscopically identify
C-rich stars even at large distances. The
differences in molecular blanketing and dust creation result in an
observed sharp dichotomy in infrared colours of O- and C-rich
stars. Thus the O- and C-stars are often identified using narrow
filters. The $J-K$ colour of the C-rich stars is $>1.4$ mag and
systematically redder than that of O-rich stars
\citep{1990ApJ...352...96F, 1996AJ....112.2607C, 2000A&AS..144..235C}.
These effects influence the luminosity function of
O- and C-rich AGB stars. The Magellanic Clouds are good
test-beds of theories of the late stages of stellar evolution.
They are nearby and yet far enough to ignore the LMC thickness, so that
all stars are approximately at the same distance.

Late evolutionary stages are often associated with long period-
(LPV), semi-, and irregular-variability. A wealth of data on this
variability was collected as a by-product of the micro\-lensing
surveys. All yielded catalogues of numerous variable stars. In these 
data the period-luminosity relations of the LPV's on the AGB are well 
documented for both the Large and Small Magellanic Clouds 
\citep{1999IAUS..191..151W, 2000PASA...17...18W, 2001A&A...377..945C, 
2003A&A...406...51C, 2002MNRAS.330..137N, 2004MNRAS.348.1120N, 
2002A&A...393..563L, 2004MNRAS.353..705I, 2004MNRAS.347..720I, 
2003MNRAS.343L..79K, 2004MNRAS.347L..83K, 2004AcA....54..129S, 
2004AcA....54..347S, 2005AcA....55..331S, 2004A&A...425..595G, 
2005AJ....129..768F, 2005A&A...438..521R}.

The period-luminosity diagram for red variables at advanced
evolutionary stages reveals six sequences for different classes of
objects \citep{2004MNRAS.347..720I}. Miras and some low-amplitude
semiregulars pulsating in the fundamental mode form the sequence
C. Other semiregular stars pulsating in the second and third
overtone modes occupy sequences A, B, and C, and the RGB eclipsing binaries form their own sequence E. The origin of
the sequence D corresponding to long-secondary periods (LSP)
remains ambiguous. \citet{2004ApJ...604..800W} considered a number of
possible explanations of LSP (radial and non-radial pulsations,
rotations, orbiting companions, chromospheric activity, orbiting
dust clouds), but none fit the observations satisfactorily. 

\citet{2004AcA....54..129S} combined the OGLE-II and
OGLE-III data and found the multiperiodicity of the red giants to be
variable with a small amplitude. They exhibited two modes closely
spaced in their power spectrum. This is likely to indicate
non-radial oscillations. They also show that members of the
short-period P--L sequences below the TRGB constitute a mixture of
the RGB and AGB variables. Recently, \citet{2004AcA....54..129S, 2005AcA....55..331S, 2007ApJ...660.1486S},
and \citet{2006ApJ...650L..55D} have demonstrated
that the sequence E overlapped with the sequence D, which may
be evidence for a binary origin of the sequence D.

There are differences between the O-rich and C-rich LPVs, too. \citet{2003A&A...406...51C}
note that C-stars have a larger amplitude than O-stars. \citet{2004MNRAS.353..705I} confirm that O- and C-rich Miras
follow different period vs. $(J-K)$ colour relations \citep{1989MNRAS.241..375F}.
The I-band amplitude of C-rich Miras tends to grow with the
redder mean $(J-K)$ colour, while the amplitude of O-rich Miras
is colour independent \citep{2005MNRAS.364..117M}.

The evidence has provided new and significant constraints
for theoretical pulsation models. Models and observations are
compared in three ways: using stellar isochrones (e.g. \citealt{1996A&A...311..425B,
2003MNRAS.344.1000B, 2002A&A...394..125M, 2003A&A...403..225M, 1993A&A...267..410G}),
stellar tracks \citep{1993A&A...267..410G, 1999A&A...344..123M, 2002A&A...393..149M, 
2002A&A...393..167L, 2003MNRAS.338..572M}, and the fuel consumption theorem by 
\citet{1986seg..conf..195R} and \citet{1998MNRAS.300..872M, 2005MNRAS.362..799M}.

The synthetic characteristics of AGB stars are derived either from
complete evolution tracks, semi-empirical fits to the core
mass-luminosity relation, or from the core mass-interpulse period relation
and the mass-loss rate as a function of stellar parameters. The
examples of such calculations are provided by \citet{1993A&A...267..410G}
and \citet{2004MNRAS.350..407I}.

Recent attempts have been made to include the TP-AGB phase in
evolutionary population synthesis models. Both improvements
in the low-temperature opacities and peculiarities of the actual surface
chemical composition have such profound consequences that the whole AGB evolutionary scenario
became significantly affected \citep{2008A&A...482..883M}. Use of the molecular opacities
reflecting the actual chemical composition leads to a significant
decrease in $T_{eff}$ as soon as C/O~$> 1$ (\citealt{2008A&A...482..883M} and references therein).
This decrease is confirmed by observations of
galactic AGB stars \citep{2001A&A...369..178B, 2003A&A...403..225M},
and it naturally explains the presence of a red tail in
the $(J-K)$ colour-magnitude diagram of C stars.

The mass loss is driven by pulsation in a complicated way. The
pulsation pushes a fraction of the atmosphere above the photosphere,
creating a cool and dense environment where dust grains form and
grow efficiently. The radiation pressure on the dust combined with
the momentum carried in the shock waves drives the mass loss \citep[e.g.]{1979ApJ...227..220W, 1991ApJ...375L..53B, 1996A&A...314..204H}

\citet{2005A&A...441.1117L} compared the observations of LPV in 47
Tuc with the models. On one hand, they find that the models
without mass loss fail to reproduce the observed periods of the
small amplitude pulsators. On the other, the $K-\log P$
sequence of the large amplitude variables, such as the Miras, is
inconsistent with the mass-oss models and consistent with the no
mass-loss models. Only models involving the non-linear fundamental
mode yield periods consistent with the Miras in \mbox{47 Tuc}
\citep{2005MNRAS.362.1396O}. 

The past decade has brought results of extensive photometric surveys --
OGLE \citep{1994astro.ph.11004P}, OGLE II \citep{1997AcA....47..319U},
EROS \citep{1995A&A...301....1A}, MACHO \citep{1997ApJ...486..697A}, 
and MOA \citep{2001MNRAS.327..868B} --
covering several optical and infrared bands, sometimes simultaneously. 
They have provided a wealth of observations of red variable stars, thus enabling 
the study of their population properties. In particular, use of infrared colours 
and/or pulsation periods have enabled the classification
of C- and O-rich stars. We discussed above the relevant observational and theoretical results.

In the present paper, we attempt to analyze the photometric properties of red variables 
in the visual region without recourse to their periods. Our analysis should be complementary 
to any traditional methods while suffering less, if at all, from aliasing 
and seasonal interference. We employ correlation slopes of colour and magnitude
variability introduced by \citet{2004ApJ...604..800W}. In this paper 
the slope $a_R$ of the correlation of $R$ magnitude and $V-R$ colour variations 
revealed no relation with any other property of red variables. 
We employ a different filter combination in our attempt of photometric classification of C 
and O-rich red variable stars. Our results 
should not depend much on the number of pulsation cycles covered by a given 
survey. Our practical requirement that data span at least one pulsation cycle, 
is met in most large surveys.

In Sect. \ref{s2} we describe data
employed in the present study. Our calculation methods are described in Sect.
\ref{s3}. In Sect. \ref{s4} we propose new methods of selection O- and C-rich stars. 
Due to large differences in filters and time span, we present our results for
each survey separately. We give an estimate of the C/O ratio for variable 
stars in the Magellanic Clouds. 
In Sect. \ref{s6} we employ a simple model to demonstrate how observed effects may arise.
Possible peculiar variables are discussed in Sect. \ref{s7}. 
We conclude in Sect. \ref{s8}.

\section{The data}\label{s2}

Our paper is based primarily on the EROS-2 photometric survey of 
the Magellanic Clouds cross-referenced with the 2MASS infrared magnitudes
(Sect. \ref{s21}).
We explicitly indicate when these are supplemented by the OGLE and MACHO
photometry and four catalogues of C-and O-rich stars (Sect. \ref{s22} -- \ref{s24}).

\subsection{EROS-2 survey}\label{s21}

The Exp\'{e}rience de Recherche d'Objets Sombres (EROS-2) project employed 
extensive photometry obtained with
the 1-meter MARLY telescope at La Silla Observatory, Chile, to search for
the baryonic dark matter of the Galactic halo by means of the
gravitational microlensing \citep{2003A&A...404..145A,
2000A&A...355L..39L, 1998A&A...332....1P,  2007A&A...469..387T}. The
observations were performed between July 1996 and February 2003
through a dichroic plate-splitting light into two wide field
cameras covering $0.7\times 1.4^o$ in right ascension and
declination each, yielding two broad passbands. The so-called blue
channel ($420-720$ nm, hereafter $B_E$ band) overlapped the standard
V and R standard bands, while the red one ($620-920$ nm, hereafter
$R_E$ band) roughly matched the standard I filter. Each camera
constituted a mosaic of eight $2k\times 2k$ CCDs with a pixel size
on the sky of $(0.6")^2$.

Ten fields covered the SMC and 88 fields covered the LMC.
%\citep{2002A&A...389..149D}.
The photometry from individual images was combined into light
curves using the Peida package developed
specifically for the EROS experiment \citep{1996VA.....40..519A}. The estimated
accuracy of this photometry is discussed by \citet{2002A&A...389..149D}.
For uniformity, the SMC data were analysed again with the more recent version of Peida.

The RGB and AGB variable stars were selected from the SMC and LMC lists of EROS variables using
their location in the colour-magnitude diagram. In this way we
selected 28914 stars in LMC and 5930 in SMC. They constitute our
sample used in further analysis.

\subsection{OGLE data}\label{s22}

The Optical Gravitational Lensing Experiment (OGLE-II) data were
collected with the 1.3 m Warsaw Telescope at the Las Campanas
Observatory, Chile, operated by the Carnegie Institution of
Washington. The $I$-band data span about 3000 days: from January
1997 to April 2005. The $V$-band measurements were obtained
from 1997 to 2001 \citep{2001AcA....51..317Z}, so they span a
shorter time baseline. Up to 70 $V$-band points per star are
available. In the $I$-band, 500 to 900 measurements were available,
depending on the field \citep{2005AcA....55..331S}. In the present work we
use data on 3586 LPV from LMC published by \citet{2005AcA....55..331S},
downloaded from the OGLE homepage\footnote{http://sirius.astrouw.edu.pl/$^{\sim}$ogle/}.

We cross-identified EROS and OGLE objects within a
$1.3"$ radius. The OGLE-II fields cover only the bar of LMC, i.e. a much
smaller field than for EROS, hence there are relatively few stars in common.

\subsection{MACHO data}\label{s23}

The MAssive Compact Halo Objects (MACHO) project \citep{1997ApJ...486..697A} comprises eight years of
observations of the Large and Small Magellanic Clouds and of the
Galactic Bulge. For the present purposes we selected LMC stars from
the MACHO catalogue of variable stars \citep{2003yCat.2247....0A}.
Using the MACHO\footnote{http://wwwmacho.mcmaster.ca/} interface, we downloaded 2868 individual light
curves of the stars classified as the red giant variables
(classified as Wood A, B, C, and D sequences).

\subsection{2MASS survey data.}\label{s24}

The Two Micron All Sky Survey (2MASS)\footnote{http://www.ipac.caltech.edu/2mass/}
employed two 1.3-m robotic telescopes, one at Mt. Hopkins, Arizona, and one at CTIO, Chile. Each
telescope was equipped with a three-channel camera,  capable of
observing the sky simultaneously at $J$ (1.25 microns), $H$ (1.65
microns), and $K_s$ (2.17 microns). The southern facility began
collecting Survey data in March 1998 and conducted its final scan
in February 2001.

We extracted 2MASS data for each of the EROS fields separately.
Then cross-identification was done separately for each field.
During the cross-identification of EROS and 2MASS, we employed a $3"$ 
search radius. We selected only stars with photometry available in
all three bands $J$,$H$, and $K_s$. AGB stars are among the brightest stars of the LMC. 
If an area of the order of 16 square degrees contains
25000 AGB stars, their average separation is 1.5 minutes of arc, 
so the probability of blending two of them within the search radius is $(1.5*60/3){^{-2}}=0.0006$, 
thus corresponding to fewer than 15 stars in the whole sample. 
This is negligible compared to the blur of our photometric plots. 
At this point it is worth discussing the contamination of our AGB sample
of IR colours with mistaken cross-identification with  RGB stars. For stars brighter than TRGB ($K<12$ for LMC and $K<12.7$ for SMC),
this problem does not exist. Suppose, for an EROS AGB star, that
IR colours for an RGB one were fitted, then the star would
land leftwards of the vertical lines on $a_V$ versus $K$ diagrams.
One possible problem is mixing faint AGBs with RGBs, causing pollution
of RGBs with AGBs but not otherwise [i.e. faint AGB remain clean,
if possibly incomplete]. \citet{2004MNRAS.347L..83K} consider the same problem in more detail for stars from OGLE and 2MASS and conclude that contamination is insignificant, possibly no more than 0.3\%, for their search radius of 1",
corresponding to less than 3\% contamination for our radius of 3".
Since there were selected MACHO objects than for
EROS, we cross-identified MACHO and 2MASS
without subdivision into fields and with the same $3"$ radius.
Again we only selected stars with photometry in all three bands
$J$,$H$, and $K_s$.

\subsection{The catalogues of C-rich stars}\label{s25}

We selected 1707 C-rich stars from the SMC
identified by \citet{1993A&AS...97..603R} in the low-resolution
spectroscopic survey employing the ESO 3.6 m telescope and 1185 stars
identified in the Siding Spring Observatory survey by \citet{1995A&AS..113..539M}.

A catalogue of 7760 C-rich stars in the LMC was
presented by \citet{2001A&A...369..932K}. These stars were identified
during a systematic survey of the objective-prism plates taken
with the UK 1.2 m Schmidt Telescope. To these we added the list of
C-rich stars from \citet{2004A&A...425..595G}.

\citet{2005AcA....55..331S} demonstrate a new method of distinguishing
between O-rich and C-rich Miras, SRVs and stars with long
secondary periods, relying on their $V$ and $I$-band photometry
and periods. The list of stars selected from the LMC in this way was
downloaded from the OGLE homepage. All C-rich stars from these catalogues were cross-identified
with the EROS and MACHO stars within a $3"$ radius.

\section{Data pre-processing}\label{s3}

The analysis outlined in this paper was performed separately for EROS LMC and SMC,
OGLE LMC and MACHO LMC data. We used data from all
surveys to study a path in the colour-magnitude diagram traversed by 
each variable star. We investigated correlation of the average slope of the path
with the chemical composition of the objects.
The results of this analysis are summarised in diagrams described in the
Sect. \ref{s4}.

The EROS survey produced simultaneous light curves in two spectral bands. By
applying time filters to each band light curve we obtained
a smooth, low-pass filtered light curve and its complementary high-pass light curve.
The sum of the two reproduces the original light curve. In this way for each 
spectral band we obtained 3 light curves: raw, low-, and high-pass, six in total. In further analysis we study 
correlations of stellar properties with the properties derived from these light curves. 

\subsection{Photometric calibrations.}\label{s31}

Both EROS and MACHO obtained simultaneous expositions in two bands
termed ``red'' ($R_E$ and $r$, respectively) and ``blue'' ($V_E$ and $v$, respectively) for nearly all observations. We simply
ignored observations made in just one filter.

Following \citet{2007A&A...469..387T}, we
converted the raw photometry into the standard Kron-Cousins system
using the following relations:
\begin{eqnarray}
V_{EROS} &=& 1.666 V_E - 0.666 R_E\\
I_{EROS} &=& R_E.
\end{eqnarray}

The MACHO data from the web database are available in the
raw instrumental system. Following \citet{1997ApJ...486..697A}, we
converted the raw photometry into the standard Kron-Cousins system
using the following relations:
\begin{eqnarray}
V_{MACHO} &=& v + 23.699 - 0.1804 (v - r)\\
R_{MACHO} &=& r + 23.412 - 0.1825 (v - r)
\end{eqnarray}
where $v$ and $r$ are instrumental magnitudes and $V_{MACHO}$ and
$R_{MACHO}$ are in the Kron-Cousins standard system. These
calibration formulae are estimated to have an overall absolute
accuracy of $\pm 0.10$ mag in $V_{MACHO}$ or $R_{MACHO}$ and $\pm
0.04$ mag in $(V-R)_{MACHO}$ \citep{2001ApJ...552..289A}.

The OGLE survey employs one camera, so observations in
different filters are not simultaneous. Observations in $V$ filter
were obtained much less frequently than in $I$. For each $V$
observation, we searched the nearest $I$ observations. If two $I$
points before and after a $V$ observation separated by no more
than 2 days, we performed a linear interpolation at the time of $V$
observation. For just one $I$ point nearby, no more than one day
from the $V$ observation, we assume they were measured nearly
simultaneously. In this way for most $V$ observations we found its
corresponding $I$ magnitudes. Other $V$ observations were ignored.
Because we only analyse stars with long periods, this procedure proved
reliable for our purposes.

\begin{figure*}
\centering
\includegraphics[angle=-90,width=17cm]{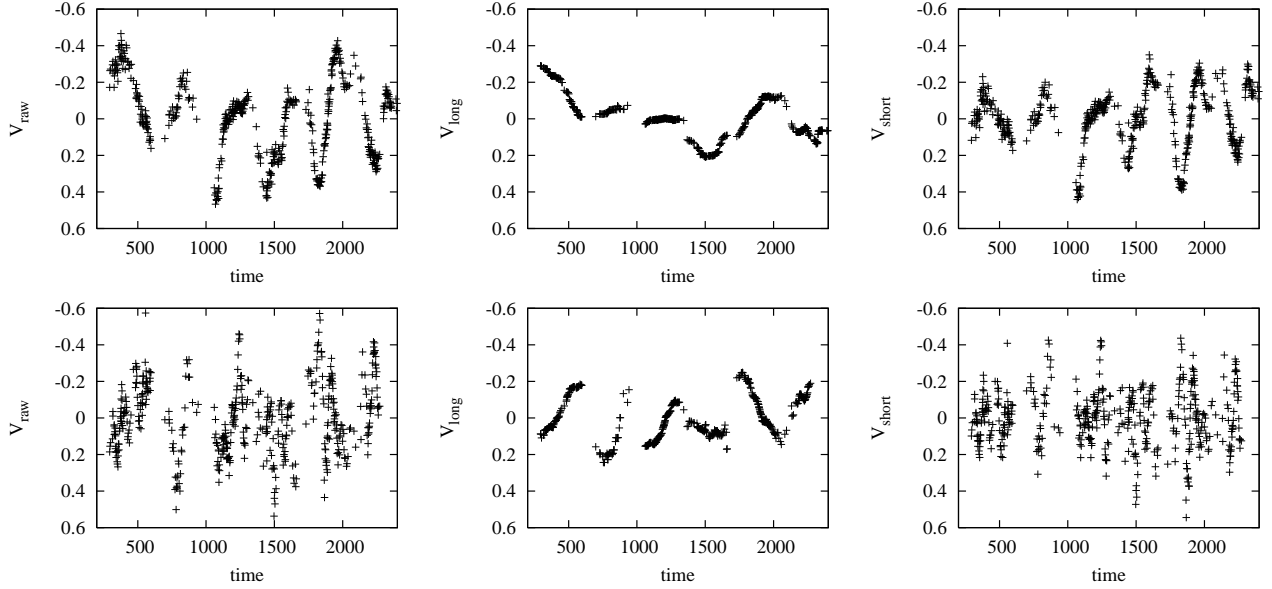}
\caption{Sample EROS light curves for two stars with spectral types assigned by 
\citet{2001A&A...377..945C}: DCMC J052618.12-694100.0 of type C (top row) and DCMC J052446.91-694949.9
of type M (bottom row). The left, middle, and right panels present the raw, low-pass, and high-pass 
light curves, respectively. For details of filtering see Sect. \ref{s32}}
\label{f0a}
\end{figure*}

\subsection{Time filtering}\label{s32}

Through the whole paper except for
Sect. \ref{s7}, we used the raw light curves.
Light curves for Sect. \ref{s7} were filtered in the time domain. The low-pass curve was obtained
from the raw one by means of the ``moving
median''. The purpose of time filtering is to separate
short period pulsations in some stars from the LSP.
Since our procedure does not depend on actual detection of these periods, 
we applied it to all
EROS red variables. We selected a window of
width $w$ centred on a given point and replaced its value with the window median value. 
In practice we filtered each light curve twice consecutively using windows  of width $w_1$
and $w_2$ dependent on average $K_s$ luminosity:
\begin{eqnarray}
  w_1 &=& 10^{(K-D)/(-3.9)}\\
  w_2 &=& 10^{(K-D+1)/(-3.9)}
\end{eqnarray}
\noindent where the $D$ magnitude is equal to 19.4 for the LMC and to 21.1 for the SMC. 
These windows were selected to have intermediate length between the short period and LSP
(see Sect. \ref{s7} for discussion of these periods).
As both periods depend on mean stellar brightness $K$, so do our window widths. 
The slope of $-3.9$ corresponds to a border line between the short period sequences and the 
LSP sequence in the period-luminosity $(P-L)$ diagram \citep{2004MNRAS.347..720I,2004AcA....54..129S}. 
As this area is virtually devoid of stars, its exact value is of no consequence for the current considerations.

We found that a light curve smoothed with two moving medians reveal LSP more clearly
than when smoothed only once.
Any short-period variations are filtered out. The high pass curve was 
obtained by subtracting the low pass one from the raw data. Only short time variations are left in 
the high-pass light curve. Sample light curves are plotted in Fig. \ref{f0a}. A similar procedure was employed by
\citet{2004ApJ...604..800W} except that we did not use the short period as
the parameter determining filter window to avoid confusion for multi-periodic/irregular
red giants \citep{2004AcA....54..129S}.
For EROS and MACHO colours $(V-I)_{EROS}$, $(V-R)_{MACHO}$ were
calculated for each time point. Subsequently we
filtered colours similarly to how magnitudes are filtered. In this way we got
the colour variation for raw data, LSP, and short time
scale variations separately. For OGLE the number of $V$ observations
was too small to filter. 

\begin{figure}
\resizebox{\hsize}{!}{\includegraphics[angle=-90]{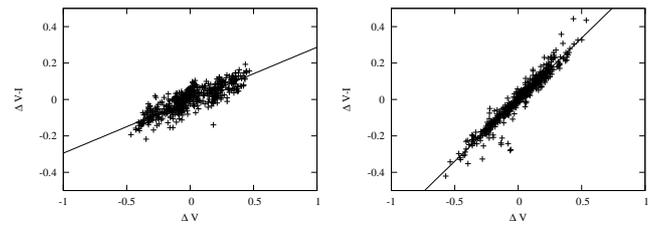}}
\caption{Sample plot of raw colour and magnitude variations for the same 2 stars as 
in Fig. \ref{f0a}. We also plot the regression lines. Their inclination parameter is $a_V$, discussed in the text.}
\label{f0b}
\end{figure}

\subsection{Analysis of colour variations}\label{s33}

In further analyses we considered variations in colour against
magnitude, for all possible combinations of light curves (raw,
long-, and short-time scale colours against raw, long- and
short-time scale $V$ and $I$ magnitudes). Sample plots are presented in
Fig. \ref{f0b}. Next we fitted by the least
squares the regression lines,
\begin{eqnarray}
  (V-I) &=& a_{I}I + b_{I}\\
  (V-I) &=& a_{V}V + b_{V}
\end{eqnarray}
\noindent where $a_V$ and $a_I$ denote the slope of the corresponding colour - magnitude
relation. This kind of linear equation was fitted for EROS, OGLE
and MACHO data.

To verify the quality of our slope parameter, we calculated the correlation coefficient $\rho$ and 
the statistic $t$ in the following way:
\begin{eqnarray}
  \rho &=& \frac{Cov\{V,V-I\}}{\sigma_{V}\sigma_{V-I}} \label{e4.1}\\
  t &=& \frac{\rho}{\sqrt{1-\rho^2}}\sqrt{n-2} \mbox{~~~where}\label{e4.2}\\
  Cov\{V,V-I\} &=& \frac{1}{n}\sum_{i}^{n}(V-\overline{V})[V-I-\overline{(V-I)}] \label{e4.0}.
\end{eqnarray}

\begin{figure}
\resizebox{\hsize}{!}{\includegraphics[angle=-90]{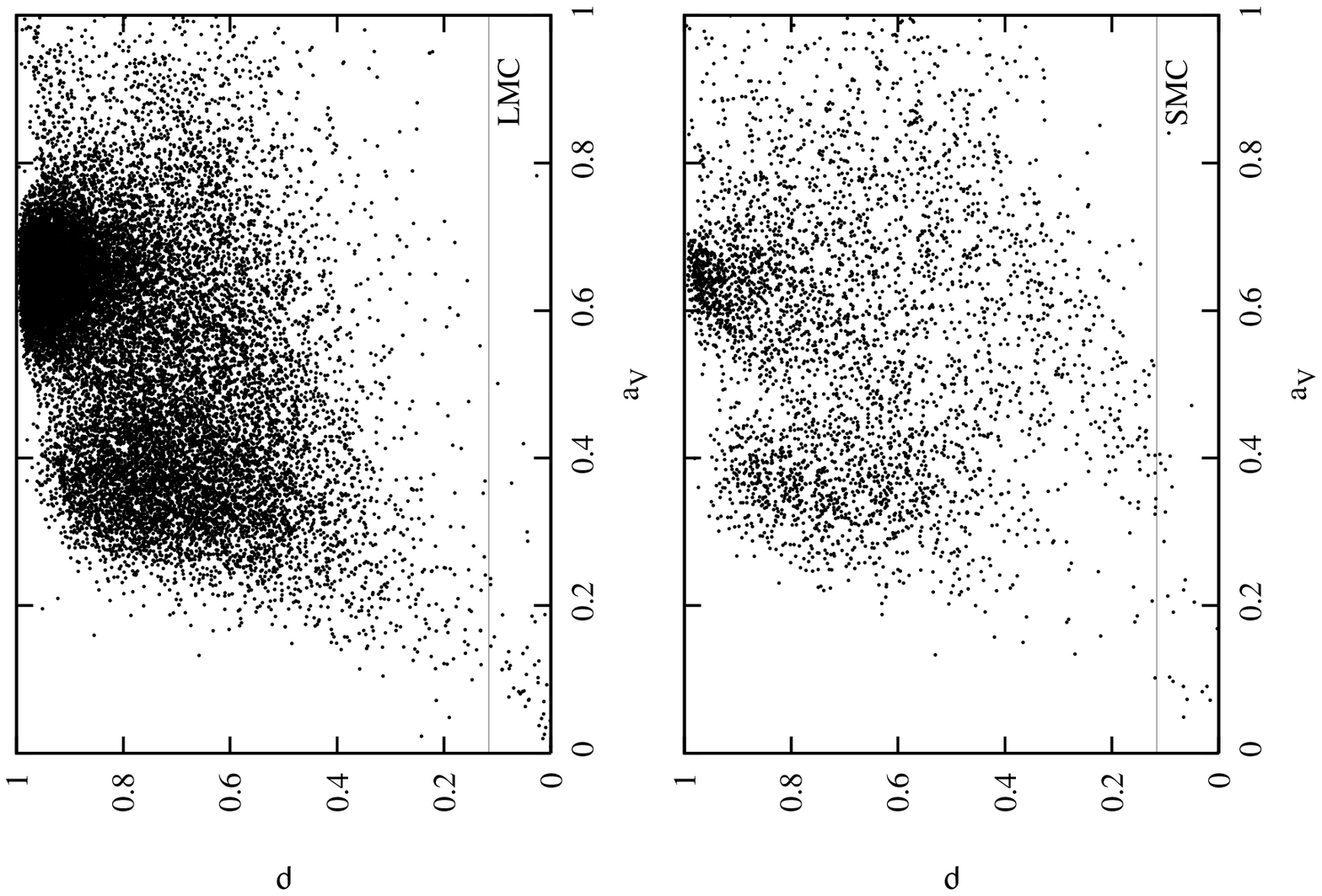}}
\caption{Plot of the colour-magnitude correlation coefficient against 
the slope of magnitude-colour relation, for EROS LMC and SMC data. 
The correlation coefficient serves as a 
measure of quality of the slope $a_v$. The horizontal lines correspond to probability 
0.995 of the null hypothesis $H_0$, i.e. points lying above correspond to statistically 
significant correlation.}
\label{f9}
\end{figure}

\noindent Results are shown in Fig \ref{f9}.
For the null hypothesis $H_0$ assuming no correlation, i.e. $\rho$ = 0, a $t$ statistic 
obeys the Student distribution with $n-2$ degrees of freedom, where $n$ is number of points 
per star \citep{fisz1963}. For a probability of 0.995, $t$ is equal to 2.6. For most stars the
correlation coefficient is greater than 0.5, and $t$ is as significant as 15 or more.

\subsection{Analysis of amplitude variations}\label{s34}

The amplitude of the luminosity variation was calculated from the
distance of the maximum and minimum in the corresponding, possibly
filtered light curve. We derived the amplitude of the colour
variation as the distance between points on these lines corresponding
to extreme magnitudes. In this way we diminished the influence of
individual colour errors on the colour amplitude. These amplitudes were
calculated for each filter, and raw , long-, and short time-scale data.

\section{Photometric chemical classification of red variable stars}\label{s4}

\subsection{The slope-amplitude diagram.}\label{s41}

\begin{figure*}
\centering
\includegraphics[angle=-90,width=17cm]{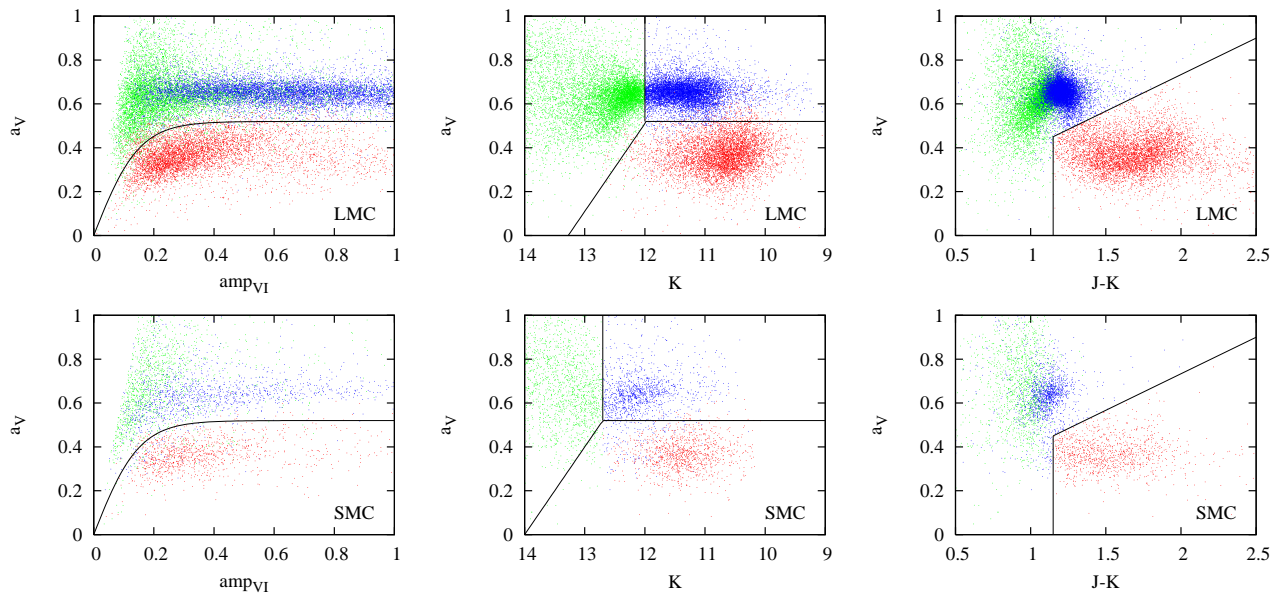}
\caption{Plots depicting three methods of distinguishing O- and C-rich variable stars (from left to right): 
slope-colour amplitude $a_{V}(amp_{VI})$, slope-$K$
magnitude $a_{V}(K)$ and slope-$J-K$ colour $a_{V}(J-K)$, for EROS LMC (top) and SMC (bottom) data. Vertical lines indicate 
adopted TRGB dividing RGB and AGB stars. See text for a detailed description of the methods and 
parameters of the boundaries. Black, dark grey, and light grey dots (blue, red, and green in electronic edition)
respectively mark O- and C-rich AGB stars and O-rich RGB ones, classified using K magnitudes for AGB/RGB
and leftmost panels for O/C content.}
\label{f1}
\end{figure*}

\citet{2004ApJ...604..800W} have introduced the variability slope parameter $a_R$. 
They note it has no obvious relation to any other property of red variable stars.
Noticing that red giants have greater amplitude in the blue filter results in
better accuracy for the slope estimates. We adopted the $a_{V}$ slope parameter
as a better substitute.  In the left panels of fig. \ref{f1} we plot the slope $a_{V}$ of the colour - blue magnitude 
relation against amplitude, for all EROS red variable stars in the LMC and SMC.
For completness we also plot $a_{V}$ against 
the near infrared magnitude and colour in the middle and right panels. In the rest of the present section, we investigate our 
slope parameter $a_V$ as a tool for the chemical and/or evolutionary status classification
of red variable stars. We defer a discussion of the classification corresponding to 
the middle and right panels until Sects. \ref{s42} and \ref{s43}.
Anticipating our result, we adopted  in Fig. \ref{f1} different
colour coding for O-rich RGB stars, O-rich, and C-rich AGB stars as green(light grey), 
blue(black), and red(dark grey) dots, respectively. The colour-coded selection, for all 
diagrams, comes from the third method presented in the right panels.

\begin{figure*}
\centering
\includegraphics[angle=-90,width=17cm]{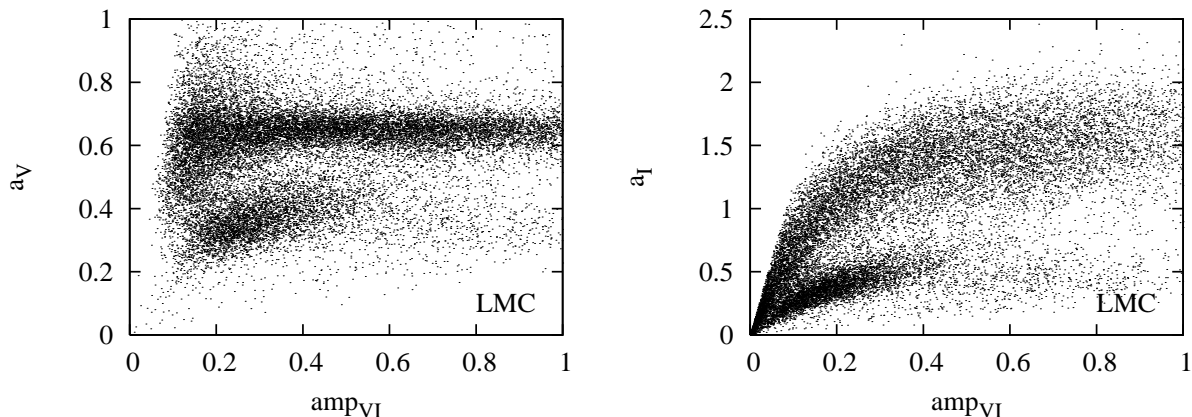}
\caption{Comparision of $a_V$ plot in Fig. \ref{f1} with that for $a_I$.}
\label{f1a}
\end{figure*}
The left panels in Fig \ref{f1} display plots of the slope $a_{V}$
against the colour amplitude $amp_{VI}$ for EROS LMC and SMC data,
respectively. Note the conspicious two-modal distribution of aV suggesting separate clustering of C- and O-rich AGB stars.
In these plots there are two distinct branches
corresponding to steep and shallow slopes, i.e. for large and small
$a_{V}$. For both branches, the slope initially grows and then, for amplitudes over 0.3,
saturates at the constant values of 0.65 and 0.3. These two groups
can be separated by the curve with the equation
\begin{equation}
  a_{V} = 0.52 \tanh(amp_{VI}/0.15) \label{e31.1}.
\end{equation}
\noindent The same border holds for both EROS LMC and SMC data.
In the following sections we demonstrate by cross-checking 
in the catalogues that the stars above and below the
curve are respectively O- and C-rich. Thus our Eq.
(\ref{e31.1}) constitutes the new classification
criterion for C- and O-rich stars. Its importance and novelty
stem from using the {\em visual} photometry alone with no recourse
to periods. Our criterion does work for an incomplete phase coverage and even for
data spanning as small an interval as a typical Mira period. Moreover,
at this point there are no obstacles to test the usefulness
of our new method for the semi- and irregular red variables.
Problems in using $a_I$ for low-amplitude variables are illustrated in Fig. \ref{f1a}. The two clusters merge for low amplitudes.
In this way we confirm that the slope $a_V$ is indeed much more suitable for classification purposes 
than our $a_I$ and $a_R$ used by \citet{2004ApJ...604..800W}. 

In Sects. \ref{s45}, and \ref{s46}, we verify our classification against existing catalogues,
while in Sects. \ref{s42} -- \ref{s44} we discuss alternative photometric systems.

\begin{figure*}
\centering
\includegraphics[angle=-90,width=17cm]{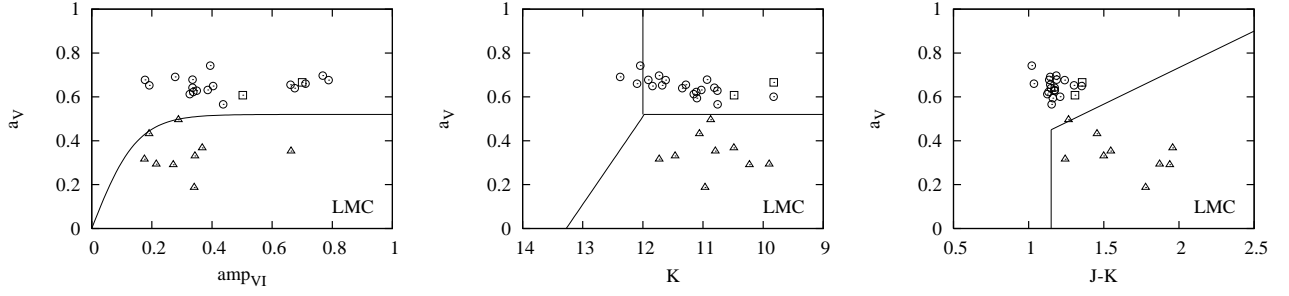}
\caption{Similar plots as in Fig. \ref{f1} for C-rich, O-rich, and intermediate stars of \citet{2001A&A...377..945C}, 
of respective spectral types C (triangles), M (circles), and S (squares).}
\label{f4a}
\end{figure*}

\subsection{Photometric properties of C-rich and O-rich stars.}\label{s45}

\begin{figure*}
\centering
\includegraphics[angle=-90,width=17cm]{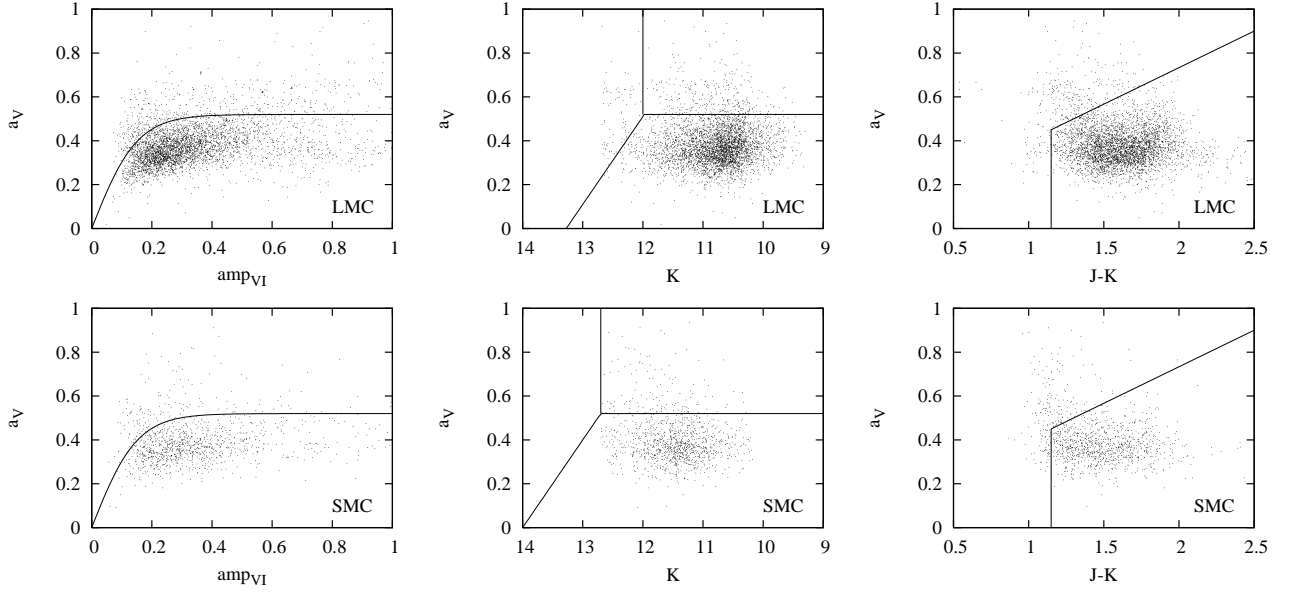}
\caption{C-rich stars in LMC (upper panels) from
\citet{2001A&A...369..932K}and \citet{2004A&A...425..595G} and in SMC (lower panels) from
\citet{1993A&AS...97..603R} and \citet{1995A&AS..113..539M}. Colour-slope
parameter $a_{V}$ in function of: amplitude of colour change  $amp_{VI}$ (left panels);
$K$ 2MASS magnitude (middle panels); $J-K$ colour (right panels).}
\label{f4}
\end{figure*}

Of the 50 stars assigned by \citet{2001A&A...377..945C} spectral types C, M, and S, 30 lie in our fields and
have entries in 2MASS. In Fig. \ref{f4a} we plot them in the same way as in Fig \ref{f1}. Spectroscopic and photometric
grouping is consistent in general and that in particular the S-stars lie on the border between C and M stars, as expected.
Clear separation of C and M stars in the left panels constitutes the most reliable test of our method.
More extensive tests follow from cross-identifying of the stars from EROS, OGLE, and MACHO with
the spectroscopic catalogues of C-rich stars by \citet{2001A&A...369..932K} and \citet{2004A&A...425..595G} for the LMC and by
\citet{1993A&AS...97..603R} and \citet{1995A&AS..113..539M} for
the SMC (Sect. \ref{s25}). For this purpose the optical ($V$, $I$) magnitudes were obtained from EROS, OGLE, and MACHO photometry,
while infrared ($J$, $K$) magnitudes were obtained by cross-reference with 2MASS  
(Sects. \ref{s31}, \ref{s22}, \ref{s23}, and \ref{s24}). In Fig. \ref{f4} we plot all identified C-rich stars
in tha same fashion as in Fig. \ref{f1}. 
The border lines in Fig. \ref{f4} are plotted for correspondence with Fig. \ref{f1}.
All previous succesful photometric criteria were based on the infrared colours 
(c.f. middle and right panels, e.g. \citet{2009AcA....59..239S}). 

Some confusion in the distribution may be caused by presence of the
intermediate S type stars with the number of O-atoms equal to that
of C-atoms. This confusion arises from the difficulty of
classifying the S stars using the low-resolution spectra obtained for
the catalogues employed in this section.

\subsection{The slope-luminosity diagram.}\label{s42}

Infrared colour and luminosity may be employed to additionally split 
red stars according to their evolutionary status.
In the middle panels in Fig. \ref{f1} we plot slope $a_{V}$ against
$K$ luminosity for EROS LMC and SMC, respectively. The separation
of the C- and O-star sequences is as clearly visible as before,
yet the O-sequence terminates abruptly near the $K$ magnitude of the TRGB
of 12 and 12.7 for the LMC and SMC, respectively \citep{2000A&A...359..601C}. 
The difference in magnitudes
corresponds to the difference in distance moduli of the LMC and SMC. The
horizontal line at $a_V$=0.52 corresponds to the criterion of Sect.
\ref{s41}. The manifestation of the TRGB here clarifies somewhat
the behaviour observed in Sect. \ref{s41}. For the O-rich RGB
stars below the TRGB the slope $a_{V}$ grows with $K$ magnitude.
As soon as they reach the AGB, the slope saturates. The skew
separation lines plotted in the figures correspond to the equations
\begin{eqnarray}
  a_{V} &<& -0.4 K + 5.31 \mbox{~~~for~~~} K>12.0 \label{e31.2}\\
  a_{V} &<& 0.52  \mbox{~~~for~~~} K\leqslant12.0
\end{eqnarray}
\noindent for the LMC and
\begin{eqnarray}
  a_{V} &<& -0.4 K + 5.51 \mbox{~~~for~~~} K>12.7 \label{e31.3}\\
  a_{V} &<& 0.52  \mbox{~~~for~~~} K\leqslant12.7
\end{eqnarray}
\noindent for the SMC. 

In the plots discussed in the present section, stars group in three clusters 
corresponding to the RGB and O- or C- AGB, but the
$K$ luminosity of the AGB and RGB stars overlaps to some extent.
A fraction of the AGB stars exists with luminosity below
that of the TRGB. Since the RGB and O-rich AGB stars have 
similar slopes, the RGB region is tainted with some O-rich AGB
stars. In the above, we only employ $K$ magnitudes for morphological purposes
to separate stars on the colour diagrams. For stars with dust shells, the
 $K$ magnitude fails as a bolometric luminosity indicator; however, in the
present paper we concentrate on optical photometry, and for the detailed discussion
of IR properties the reader is referred to the original papers.

\subsection{The slope- $J-K$ colour diagram.}\label{s43}

\begin{figure}
\resizebox{\hsize}{!}{\includegraphics[angle=-90]{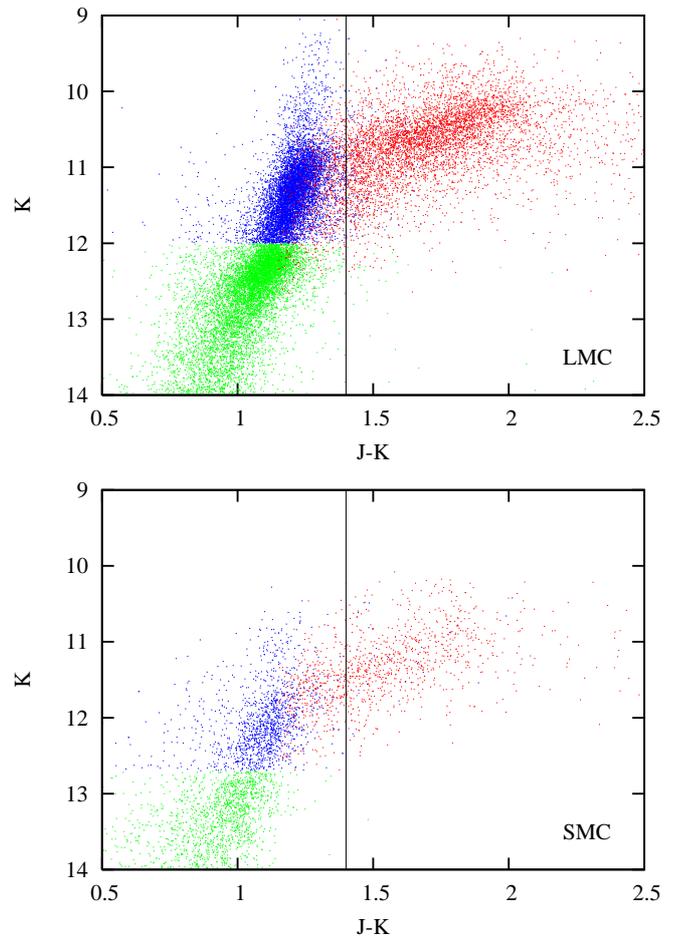}}
\caption{K - J-K diagram for LMC and SMC red variables.}
\label{f2}
\end{figure}

In the infrared, the C-rich stars appear systematically 
redder ($J-K>1.4$) than O-rich stars \citep{1990ApJ...352...96F, 
1996AJ....112.2607C, 2000A&AS..144..235C, 2005AcA....55..331S};
however, in $K$ vs. $J-K$ diagrams (Fig. \ref{f2}) there is some overlap 
between the C and O regions. The efficient way to separate the C
and O stars is to combine the slope $a_{V}$ with  the $J-K$ colour. 
In the right panels of Fig. \ref{f1} we plot slope $a_V$ against $J-K$.
Now  separate clustering of C- and O-rich stars becomes more obvious. Based on this we propose 
the following criterion for C-rich stars:
\begin{eqnarray}
  a_{V} &<& 1/3K+1/15 \mbox{~~~and~~~} J-K>1.15.
  \label{e42.3}
\end{eqnarray}
\noindent This criterion corresponds to the colour coding of stars in all panels in Fig. \ref{f1}.
Oxygen stars were separated from RGB and AGB using TRGB.

The Z-like shape of the distribution of stars in the $a_V$ - $(J-K)$ panels in Fig. \ref{f1} represent 
the evolutionary sequence from O-rich RGB (top left) through O-rich AGB (right) down to C-rich AGB objects. 
For O-rich RGB stars, amplitudes tend to be low (left panels in Fig. \ref{f1}) and 
slope parameter $a_{V}$ is poorly constrained. For higher luminosity O stars beyond TRGB, the 
slopes are better determined and stars concentrate around $a_{V}= 0.65$ and $J-K=1.2$. 
When TP-AGB begins and C-rich matter shows on the surface, the star migrates towards
the inclined border line where there are stars with intermediate MS, S, and SC types. 
Eventually after crossing the border, it lands in the C-rich AGB star cluster. There is 
no appreciable difference between diagrams for the LMC and SMC. Using $a_{V}$, $K$, and $J-K$ we cannot
only classify stars but also we could roughly determine its evolutionary stage.
No such direct evolutionary interpretation was available for our slope parameter diagrams;
however, this was made by resorting to cross identification of stars in the infrared and optical catalogues, 
and looking for which slope diagram concentrations correspond to the concentrations in the IR diagram.

\subsection{The Wesenheit index-luminosity relation}\label{s44}

\begin{figure}
\resizebox{\hsize}{!}{\includegraphics[angle=-90]{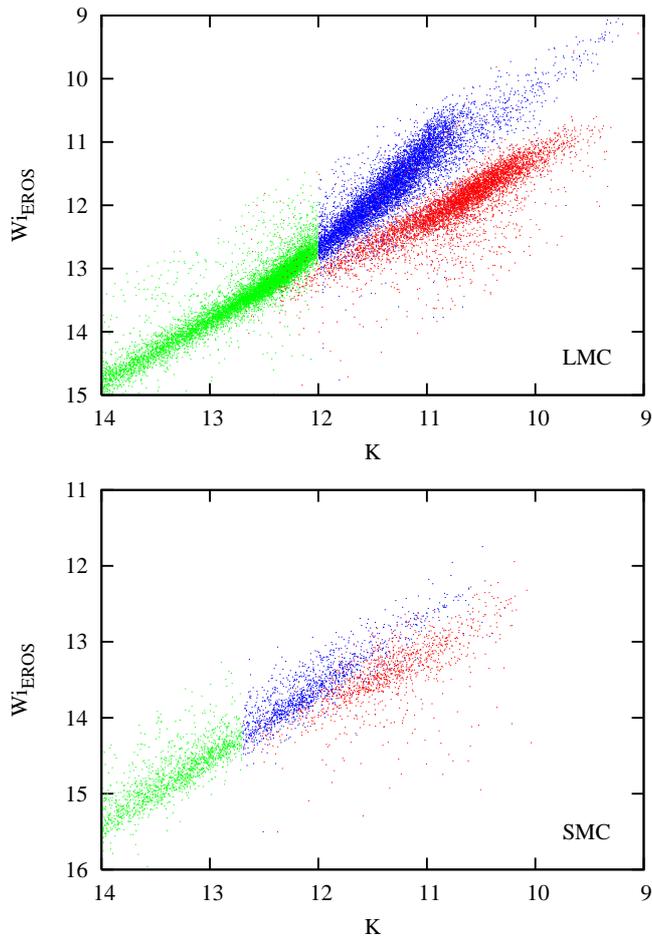}}
\caption{EROS Wesenheit index vs. K magnitude, for LMC and SMC red
variables. Wesenheit index is derived from EROS R and B magnitudes.}
\label{f3}
\end{figure}

The Wesenheit index is a reddening-free parameter
defined by a linear combination of stellar magnitudes.
For EROS bands it could be calculated from the following equation 
(e.g. \citet{1997eds..proc...91T} and references there): 
\begin{equation}
  Wi_{EROS} = I_{EROS} - 1.55 (V_{EROS} - I_{EROS}) \label{e43}
\end{equation}
\noindent where $I_{EROS}$ and $V_{EROS}$ are light curve average magnitudes (Sect. \ref{s31}).
For EROS, observations in both filters are simultaneous so time sampling
yields no big uncertainty in colours.

Plots of $Wi_{EROS}$ against the $K$ magnitude for EROS LMC and
SMC are presented in Fig. \ref{f3}. Separation of C- and O-stars 
becomes particularly clear for the LMC. The location of three groups of
stars with respect to the TRGB is visible in these plots. Similar diagrams for
the Wesenheit index were already presented by \citet{2002MNRAS.330..137N} and
\citet{2009AcA....59..239S}, but our filter bands and fields are rather different. 
In particular the accuracy of the estimation of the Wesenheit index is
sensitive to the quality of the two-band coverage. For EROS the
advantage stems from the simultaneous observation in two bands.

As the Wesenheit index is well correlated with $K$ magnitude, it is in principle 
possible to implement the method of Sect. \ref{s42} by relying purely on EROS visual 
observations. In particular the Wesenheit index would be helpful to separate the
RGB and AGB stars; however, such a classification would be less precise than 
than the one based on $K$ luminosity. The gap on TRGB, clearly visible in $K$ distribution, 
is hard to find in $Wi$. We can also separate RGB from AGB stars using the I-band. The TRGB appear at 
$I = 14.54$ in the LMC and $I = 14.95$ in the SMC \citep{2000A&A...358L...9C}.

\subsection{OGLE.}\label{s46}

\begin{figure*}
\centering
\includegraphics[angle=-90,width=17cm]{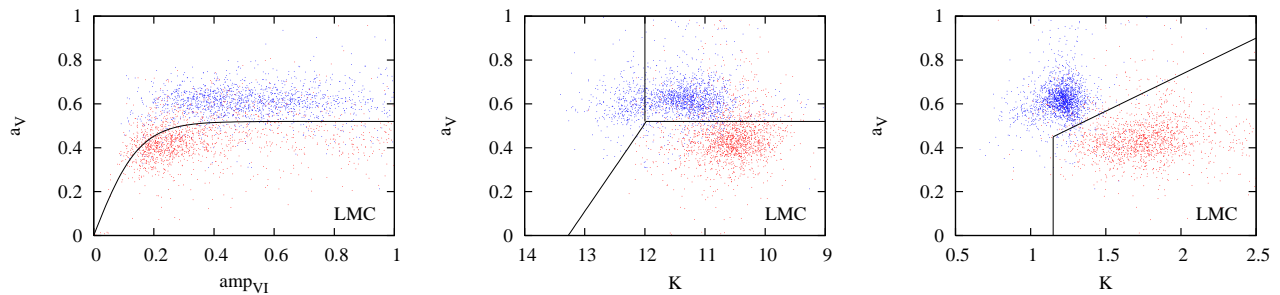}
\caption{Separation of C-rich,
O-rich AGB, and red variables below the TRGB for LMC from OGLE LPV I
and V photometry \citep{2005AcA....55..331S}. The colour slope parameter $a_{V}$ vs.
the amplitude of colour change (left panel). The colour slope
parameter $a_{V}$ vs. $K$ 2MASS magnitude (middle panel). Colour slope
parameter vs. $J-K$ colour (right panel).}
\label{f5}
\end{figure*}

The bands employed by the EROS survey were non-standard. In this
respect it is desirable to verify our results with OGLE
observations obtained in the standard $I$ and $V$ bands. For the
OGLE light curves, we performed the calculations similar to those
for EROS. In this way we obtained values of the corresponding
slope $a_{V}$. The results for OGLE are presented in Fig. \ref{f5}. We
only plot LPV selected by \citet{2005AcA....55..331S}, who reject
low-amplitude variables. Thus the number of stars with low colour
amplitude and low $K$ luminosity is lower than for EROS LMC.

Nevertheless, features revealed in Fig. \ref{f1} are present in Fig. \ref{f5}.
It is remarkable that, despite the small number of $V$ observations 
and the interpolation of $I$ magnitudes, separation of stars into
respective classes is clear.

\citet{2005AcA....55..331S} suggest that this sample should only contain the
LPV AGB stars. We found that below the TRGB their stars follow the
relation corresponding to the RGB stars. This may indicate that
these stars belong to the AGB or at least that they follow the same relation as the RGB. These authors
introduced the Period-Wesenheit index plots for a new
method to differentiate between the O- and C-rich stars. Our
method introduced in Sect. \ref{s4} is complementary as we use no information on periods. 
In this way our method is less demanding on time coverage of the
light curves. Nevertheless, the two methods of selection of O- and C-rich
stars are fairly consistent. 

\subsection{MACHO.}\label{s47}

\begin{figure}
\resizebox{\hsize}{!}{\includegraphics[angle=-90]{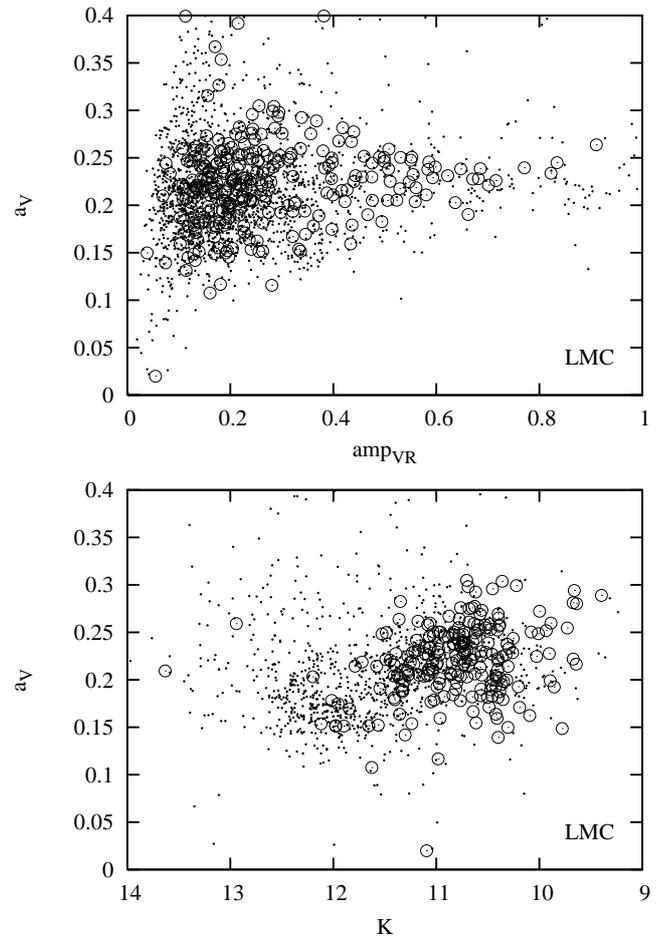}}
\caption{Slope $a_V$ against $V-R$ amplitude $amp_{VR}$ (top) and against 2MASS $K$ magnitude, 
for MACHO LPV $V$ and $R$ photometry of the LMC. 
The C-rich stars are marked with circles.}
\label{f6}
\end{figure}

The same procedure as for EROS was repeated for the MACHO data.
MACHO uses non-standard filters fairly close to the
standard $V$ and $R$ rather than $V$ and $I$. In this way MACHO bands 
differ markedly from those of EROS and OGLE.
This difference strongly affects our results for MACHO. The plots
of the slope $a_V$ against either the colour amplitude for MACHO bands or the 2MASS K mag reveal
no clear separation of the O- and C-rich stars in 
Fig. \ref{f6}. This result seems consistent with the failure of the \citet{2004ApJ...604..800W}
classification of the O- and C-rich stars based on similar filters. 
Thus our method only works for suitable photometric 
bands and is useless for MACHO photometry.

\subsection{Maps of C/O ratio for Magellanic Clouds.}\label{s49}

In Table \ref{table1} we compared the three new methods described in 
Sect. \ref{s41}, \ref{s42} and \ref{s43} with the well known method 
$K(J-K)$ that uses $J-K>1.4$ as a criterion for C-rich stars. Values 
in Table \ref{table1} were calculated only for variable stars. For all 
methods and for both Magellanic Clouds we found more C-rich stars 
than with the old method. The $K(J-K)$ method yields the conclusion that 
the C/O ratio is much lower for the SMC than for the LMC. The 
 $a_{V}(J-K)$ method yields the opposite result: the C/O ratio 
is greater in the SMC than in the LMC. Right panels of 
Fig. \ref{f1} shows that many C-rich stars lie between 1.15 and 1.4 on the
$J-K$ axis, especially for the LMC and $a_{V}(J-K)$ can separate 
stars without problem in this range of $J-K$. Different distributions of C-rich stars in 
the Magellanic Clouds probably depend on metallicity. Stars in the SMC are bluer
than in the LMC for a similar stage of evolution.

Table \ref{table1} also contains results for $a_{V}(amp_{VI})$ and 
$a_{V}(K)$ methods. To compare $a_{V}(amp_{VI})$  with other methods, we 
divided O-rich stars to RGB and AGB stars using TRGB. Both methods are affected 
by overlapping O- and C-rich regions but still we found more C-rich stars 
than with the the $K(J-K)$ method.

\begin{figure}
\resizebox{\hsize}{!}{\includegraphics[angle=-90]{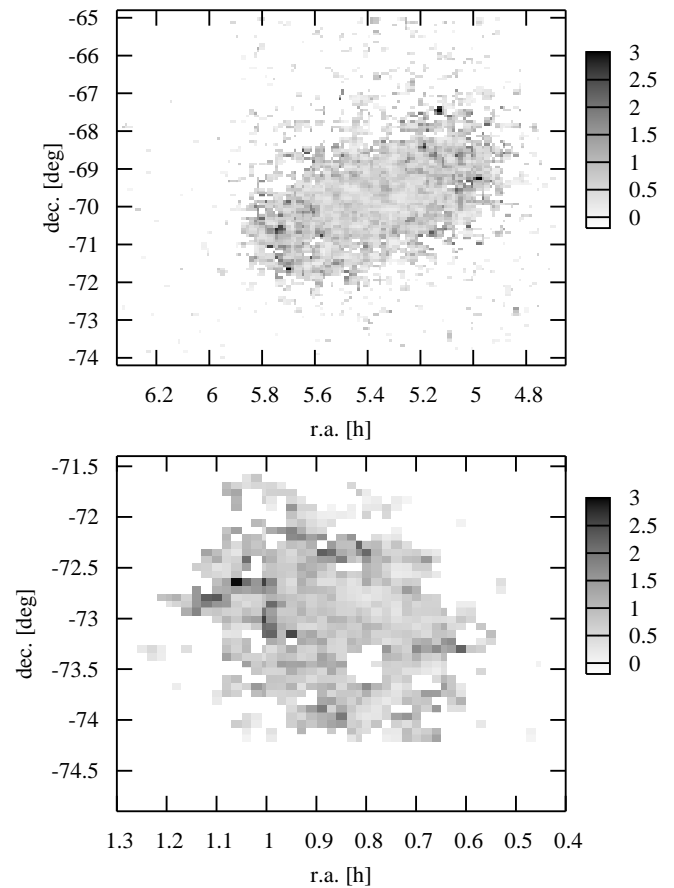}}
\caption{Maps of the C/O ratio in the LMC (upper panel) and in the SMC (lower panel) obtained
from EROS data using the $a_{V}(J-K)$ method described in Sect. \ref{s43}.}
\label{f8}
\end{figure}

From our samples of O- and C-rich AGB stars from the $a_{V}(J-K)$ method, we
calculated the C/O ratio in all cells belonging to a grid of 100' $\times$ 100'
over the face of the LMC and grid of 45' $\times$ 45' over the face of the SMC.
Figure \ref{f8} shows the ratios after boxcar average smoothing.
A large hole in the central part of the SMC results from the lack of data in that sector.
We retreive the same trends as described by \citet{2003A&A...402..133C} in 
their Figs. 3 \& 4: in the LMC the higher values of the C/O ratio are located 
in the outer regions, while the structure in the SMC is clumpier.

\begin{table}
 \caption{The C/O ratio for the MC's from variable stars}
\label{table1}
\centering
\begin{tabular}{c c c c c}
\hline\hline
Method &  $K(J-K)$ & $a_{V}(amp_{VI})$ & $a_{V}(K)$ & $a_{V}(J-K)$ \\
\hline
LMC & 0.63 & 0.69 & 0.69 & 0.72 \\
SMC & 0.46 & 0.89 & 0.98 & 0.81 \\
\hline
\end{tabular}
\end{table}

\section{Clues to the underlying scenarios}
\label{s6}
Detailed modelling of the observed effect is beyond the scope of
the present paper. As we pointed out above, modelling the late evolutionary
stages is difficult and constitutes a whole branch of the theory.
Worse, reconstruction of the synthetic spectra of C-rich AGB stars proved difficult
and strongly relies on progress in the low-temperature opacities \citep{2008A&A...482..883M}. 
Simulation of the pulsations of the AGB stars is still more
complex. The stellar atmosphere during pulses 
does not move uniformly up or down \citep{2004MNRAS.352..318I, 2004MNRAS.355..444I}.
Thus the non-linear pulsation model of a red giant is required to fit
observations \citep{2005A&A...441.1117L, 2005MNRAS.362.1396O, 2006MmSAI..77...76W}.

\begin{figure*}
\centering
\includegraphics[angle=-90,width=17cm]{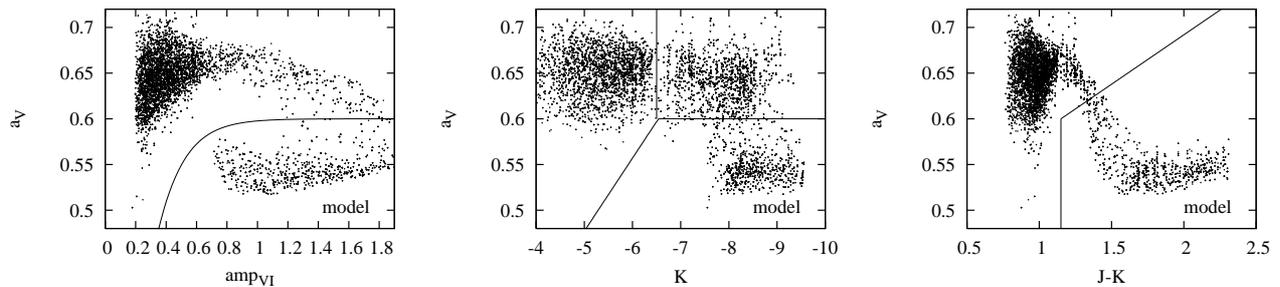}
\caption{Same as in Fig. \ref{f1} for the model slope parameter $a_V$ calculated
for real LMC stars using \citet{2003A&A...403..225M}
model sequences (see text for details). Our crude model yields a
clear separation of O-rich RGB, O-rich AGB, and C-rich AGB stars.}
\label{f7}
\end{figure*}

However, it is possible that the effect demonstrated in Sect. \ref{s4} does not depend 
strongly on the details of the underlying physics. To demonstrate that
we resorted to a crude and possibly non-physical model assuming that for all AGB stars 
during pulses $d\log L/d\log
T_{eff}=const$ and that the amplitude of variation in $L$ and
$T_{eff}$ remains the same for all stars. Thus any observed
differences between stars would follow from a different response
by their envelope to the same modulation. Next, we assume that the
response of the upper envelope is quasi-static and may be
determined by interpolation between different statics, i.e.
non-pulsating, evolutionary models from different tracks but with
the suitable $L$ and $T_{eff}$. This could be far from reality.

We employed the tracks calculated by \citet{2007ASPC..378...20G} and 
available for download\footnote{http://pleiadi.aopd.inaf.it}
to interpolate properties of individual stars at their extreme luminosities.
we take luminosity $L$, effective temperature $T_{eff}$, absolute bolometric magnitude $M_{bol}$,
absolute magnitudes in Johnson-Cousin-Glass $UBVRIJHGK$ pass-bands,
and surface carbon-to-oxygen number ratio $C/O$ listed for the tracks. 
We interpolated these values to find how luminosities in filters vary with
$L$ and $T_{eff}$ during pulsation. In each track we identified the
location of the RGB stars during TRGB and AGB phases of evolution.
Next we perturbed their $L$ and $T_{eff}$ to simulate the
pulsation extreme phases. Using the corresponding filter
magnitudes, we calculated slopes and amplitudes and plotted them in a
similar way as for Fig. \ref{f1}. By trial and error we found that
a reasonable match with Fig. \ref{f1} is obtained for $\Delta\log T_{eff}=0.042$, $\Delta\log L=0.026$,
which leads to ${d\log L}/{d\log T_{eff}}=0.614$, but this works well 
for a wide range of both parameters. In this way we obtain Fig. \ref{f7} for the simulated data, where
\ref{f7} O- and C-rich stars separate in
the slope-colour amplitude, slope-$K$ luminosity and slope-colour $J-K$ diagrams.

As our model is entirely artificial, we refrain from strong conclusions. However, this result should
encourage theoreticians to reproduce our effect using fully realistic calculations (e.g. \citealt{2008A&A...482..883M}),
and to use it to restrict the pulsation scenario.

\begin{figure}
\resizebox{\hsize}{!}{\includegraphics[angle=-90]{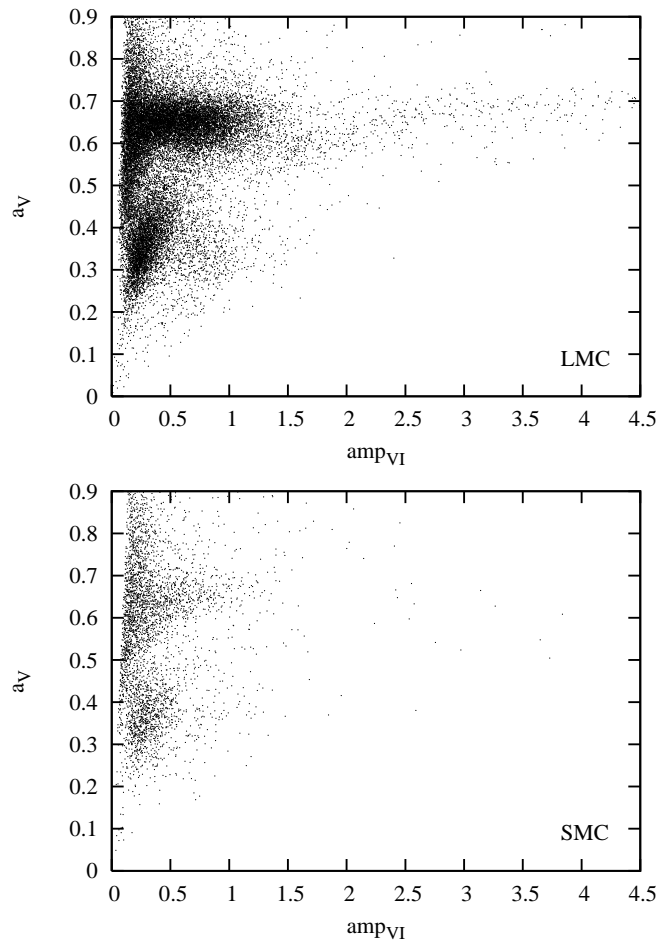}}
\caption{Colour-slope parameter in function of amplitude of colour change for EROS LMC and SMC data.
Crosses mark large-amplitude variables.}
\label{f12}
\end{figure}

\begin{figure}
\resizebox{\hsize}{!}{\includegraphics[angle=-90]{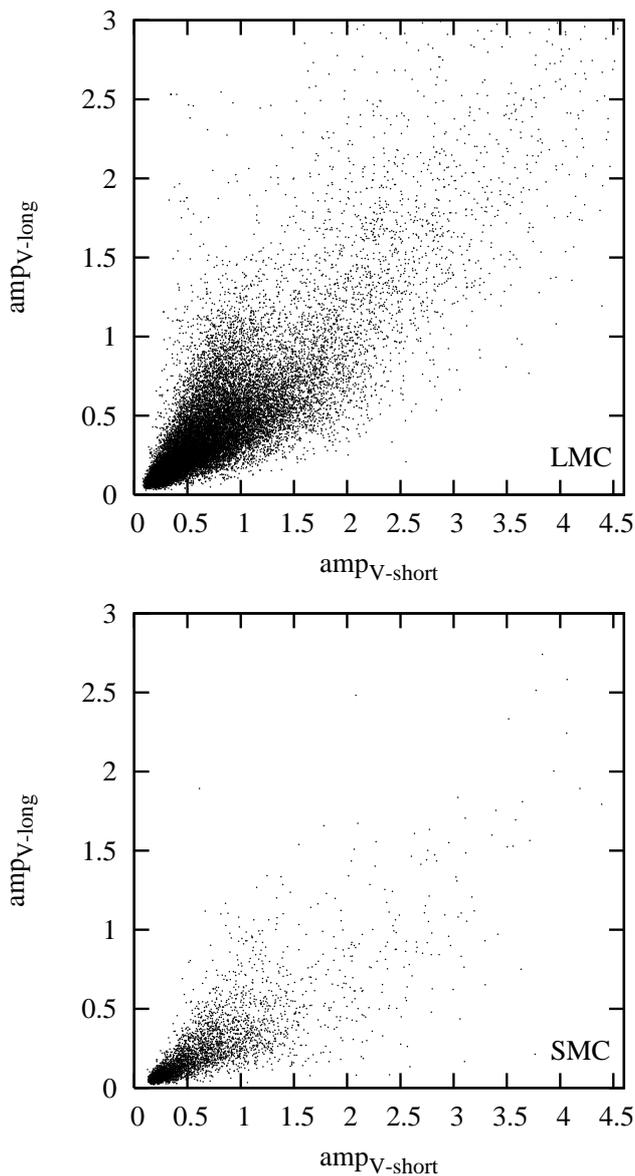}}
\caption{Correlation between amplitudes of LSP and of shorter period variation for EROS LMC and SMC data. Crosses mark large-amplitude variables.}
\label{f11}
\end{figure}

\section{Special properties of red variables.}\label{s7}

So far, our analysis has concerned the red variable stars with a modest amplitude ($amp_V<1$ mag.),
since they posed biggest classification problems. Also we payed no special attention to the multiperiodic variables
in general and those with LSP in particular. Although our method 
for these stars does not explicitly account for their peculiarites, 
we briefly present results of its application hoping to provide additional 
clues missed in more specialised analyses. 

\subsection{Large-amplitude pulsations}\label{s71}

In Fig. \ref{f12} we plot the slope- against colour
amplitude for the whole range of amplitudes. This figure, which based on the raw data,
corresponds exactly to the left column in Fig. \ref{f1}, 
except for the extended amplitude scale. To illustrate here and in the next figure
the relation of colour and luminosity amplitudes, $amp_{VI}$ and $V$, stars to the right of 
the line of inclination 0.6 passing through the origin are marked with different symbols.
It appears that the pattern of the two populations of stars with 
small and large slopes discussed in Sect. \ref{s4} does extend to the large amplitudes, too.  
On a related plot of Wesenheit index vs. $K$ luminosity, the large-amplitude
C-rich stars appear redder, while the
O-rich ones are bluer then the corresponding stars of small amplitude. 
We found no other peculiarities for the large-amplitude stars. 

\begin{figure}
\resizebox{\hsize}{!}{\includegraphics[angle=-90]{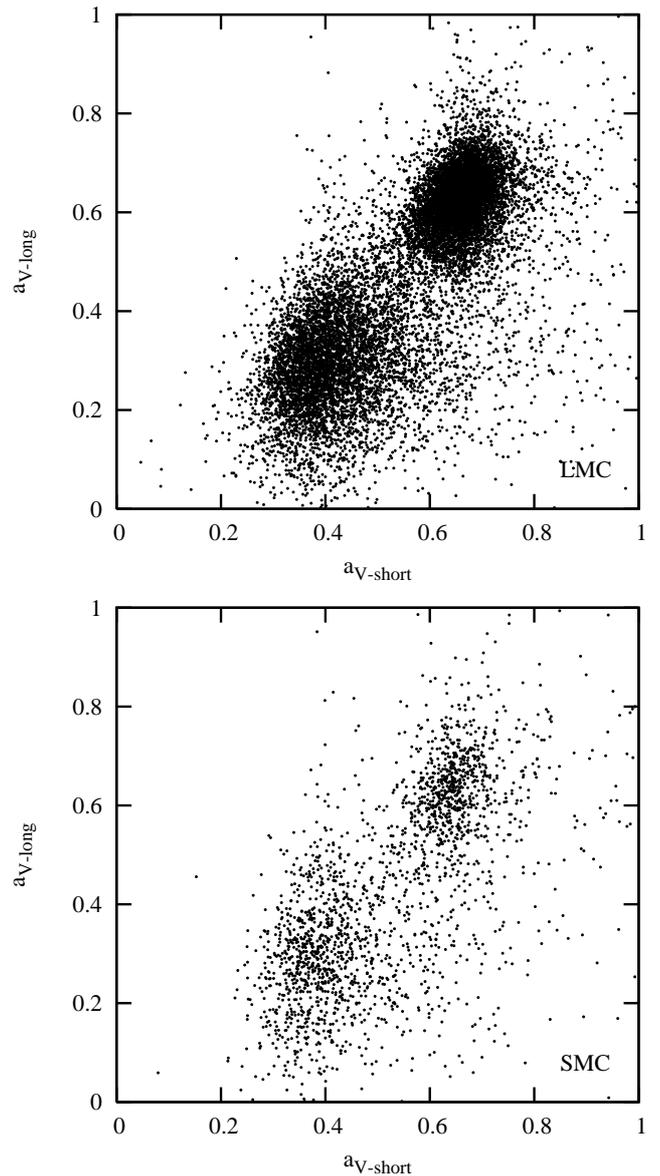}}
\caption{Correlation between the colour slope parameter of LSPs and the shorter period for EROS
LMC and SMC data.}
\label{f10}
\end{figure}

\subsection{Long secondary pulsations}\label{s72}

Up to now we have analysed the raw data. However,
many RGB variables reveal combinations of two periodic variations:
a short one, presumably a radial pulsation, and a LSP.
The corresponding periods form sequences $B/B'$ and $D$ in the period-
luminosity diagram (P-L). The nature of the LSP in RGB stars is still
a mystery for interpretation (\citet{2007ApJ...660.1486S,  2009MNRAS.399.2063N} and
references therein).
The $D$ sequence periods are roughly ten times longer than the $B$ ones 
and correspond to no known radial pulsation mode. The $B$ sequence may correspond to a low-number overtone
radial mode. LSP is observed among at least 25\% of low mass RGB stars and typical red amplitudes
$amp_R$ do not exceed 0.8 magnitudes.

In this section we apply methods of Sect. \ref{s4} to the time-filtered light curves of 
{\em all} EROS LMC and SMC red variable stars, whether known to display an LSP or not. 
In the process we use the fact that our procedure does not require a known period. 
By employing the filters described in Sect. \ref{s3} we
extracted separate light curves for fast and slow variations. 
The cut-off frequency of our filters depends on the mean magnitude 
to ensure a proper split of $B$ and $D$ sequences in systems known to display LSPs.
Thus we obtain two different sets of amplitude and slope parameters 
$amp_{V-short}$ and $amp_{V-long}$, $a_{V-short}$ and $a_{V-long}$ from respectively 
high- and low-pass filtered light curves. 

In Fig. \ref{f11} we plot long-time amplitudes of variation of magnitude against short time ones, 
$amp_{V-long}$ v.s. $amp_{V-short}$ for all EROS LMC and SMC data. Stars marked
in Fig. \ref{f12} as possessing {\em relatively} large colour magnitude $amp_{VI}$ also exhibit {\em absolutely} large
magnitudes $amp_V$ on at least one time scale. Although the
amplitudes in Fig. \ref{f11} were estimated independently from two different light curves, they seem correlated.
A consistent yet non-unique explanation would involve a similar mechanism for the short and long
light variations. 

A stronger argument supporting the similarity of mechanisms of short and long time variations 
stems from the tight correlation of the short and long time scale slopes $a_{short}$ and $a_{long}$
revealed in Fig. \ref{f10}. As discussed in Sect. \ref{s4} the two separate blobs in Fig. \ref{f10} 
correspond to C- and O- rich stars. The presence of just two such blobs rather than three or four at square vortices
provides striking demonstration that short and long time scale variations
both depend in the same way on surface chemical properties. Combined evidence from
Figs. \ref{f11} and \ref{f10} indicates that both
long and short time scale variations are due to the reaction of the stellar photosphere to 
similar kind of perturbations. We caution to not overinterpret this result.
Firstly, our data contain both stars with and without known LSPs, and most light curves
exhibit both periodic and irregular variations. However, it must be remembered
that LSP amplitudes are not that small, and they are present in a sizable fraction of stars.
If the LSP were produced away from the photosphere, one could expect to see some substructure
within both C and O blobs in Fig. \ref{f10}, so apparent lack of any internal 
structure in these blobs  is consistent with LSPs being caused by
the same kind of photosphere response as for the irregular variations and for the short period
pulsations. In this context we stress that the EROS two-band observations are strictly simultaneous;
hence our conclusions on photosphere evolution involve no interpolation or implicit 
light curve models.

\section{Conclusions.}\label{s8}

A variable star during each cycle completes a closed loop on the colour-magnitude diagram.
Non-periodic variables travel more involved routes. For each red variable star from the EROS survey,
we derived a width and average slope of these figures, respectively $amp_{VI}$ and $a_V$.
In doing so, we benefitted from reliable colours obtained from EROS simultaneous 
two-band observations. In this way our analysis is less affected by variability and/or 
time sampling patterns. In particular, the gaps and 
aliasing pose no big problem for us as long as the data cover most of the pulsation period.
This is a much less demanding requirement than a span of several cycles needed for
reliable period analysis. The price paid in the process is a loss of any extra information;
nevertheless, our analysis yields independent results that supplement more traditional period studies.

Our key method derived in Sect. \ref{s41} relies on the parameters $a_V$ and $amp_{VI}$
derived from pure visual observations. For its testing and verification
we cross-referenced stellar positions from the EROS, OGLE and MACHO optical surveys with 
the infrared 2MASS catalogue. $K$ magnitudes from the last catalogue were treated
as indicators of stellar mean luminosity. Such a procedure is reliable for a known distance 
because it has little  $K$ band extinction and small bolometric correction for red variables.
To completely free our optical luminosities from extinction, we employed the Wesenheit index. 
However, our key parameters result $a_V$ and $amp_{VI}$ are independent of distance and magnitude
values. The range and correlation of two band magnitude changes are the only parameters that are important for us.

We demonstrated that in the $a_V$ and $amp_{VI}$ diagram, red variable stars form two
well defined bands along $a_V=0.4$ and $0.65$ lines. The bands occupied the same location 
for both the LMC and SMC stars. We argued that such a diagram constitutes a new method for 
differentiating of C- and O-rich RGB stars from purely visual observations. So far,
effective photometric methods have relied on infrared observations. In fact,
the dichotomy of C/O star distribution becomes even more prominent in the 
slope $a_V$- K luminosity and slope- (J-K) colour diagrams for all stars from our sample.
The clustering is particularly pronounced, both for LMC and SMC, in the slope-luminosity
diagram. There the AGB and RGB stars separate well, apart from the usual C/O split of $a_V$. 
In this particular diagram stellar evolutionary tracks RGB-TRGB-AGB(O)-AGB(C) are 
particularly well defined. While the detailed location of clusters in our diagnostic diagrams
does depend on passbands, it is clearly manifested both for both EROS and OGLE two-band photometries.
However, the bands employed by MACHO are not suitable to our classification method. 

The final proof of the physical meaning of the clustering observed in our diagrams 
stems from the cross-check with catalogues of C-rich stars.  The overwhelming 
majority of C-rich stars  from catalogues by \citet{2001A&A...369..932K}, 
\citet{2004A&A...425..595G}, \citet{1993A&AS...97..603R}, \citet{1995A&AS..113..539M},
for LMC and SMC, belongs to the low- $a_V$ slope cluster.
Thus it appears that indeed the C- and O-rich stars belong to separate clusters 
in our $a_V$ vs. $amp_{VI}$ diagnostic diagram.
In this way we discovered and verified a new method of distinguishing between C- and O-rich
stars. Our method works for sparse coverage in two-colour light
curves in the bands resembling $V$ and $I$. Since we rely on
colours, the measurements in both filters should be (nearly)
simultaneous. Our method was tested successfully for EROS-2 LMC
and SMC observations and also for OGLE data. It does not seem 
to suffer from any metallicity effect as cluster locations for the LMC and SMC
were identical. Using simple simulations based on realistic evolutionary models of RGB,
we demonstrated plausible cause of the effect revealed by our method;
however, for as broad filters as those used by MACHO our method fails. One
future application of our diagrams is to test the pulsation evolutionary models
on a large population of RGB and AGB stars.

Using our method of differentiating the O- and C-rich stars we were able
to derive the population C/O star ratio, an indicator of the
mean metallicity of a stellar population, as well as a tracer of the
history of stellar formation \citep{2006A&A...448...77C}. Our selection
can recognise more C-rich stars than using $J-K>1.4$ criterion, and it yields
higher values of C/O ratio. Thus we were able to produce C/O maps for the LMC and SMC.

We payed some attention to colour correlation properties of the LSP compared to the short primary pulsation (SPP).
Our results provide some evidence favouring origin of both LSP and SPP
in the stellar photosphere. This follows from the correlation of
colour-magnitude slopes and separately from correlation of amplitudes 
for the LSP and SPP. In other words, both SPP and LSP depend on the chemical (C/O)
properties of the photosphere. The colour variations would
be consistent with modulation of the effective temperature.
This seems to exclude any aspect effects, e.g. eclipses, as a cause of LSP;
however, using our purely photometric observations it would
be premature to conclude that both SPP and LSP clocks are due to pulsation.
However, both clocks should somehow affect the photosphere in similar ways.

\begin{acknowledgements}

We would like to thank Jim Rich for his careful reading of the 
manuscript and the anonymous referee for helpful comments.
This work made use of EROS-2 data, which were kindly provided by the EROS 
collaboration. The EROS (Exp\'erience de Recherche d'Objets Sombres) project was 
funded by the CEA and the IN2P3 and INSU CNRS institutes.
We acknowledge use of data the Two Micron All Sky Survey,
which is a joint project of the University of Massachusetts and
the Infrared Processing and Analysis Center/California Institute
of Technology, funded by the National Aeronautics and Space
Administration and the National Science Foundation. This work was 
carried out, for ASC, JPB, JBM, and MW,
within the framework of the European Associated Laboratory 
``Astrophysics Poland-France''. ASC acknowledges support by the
Polish grant MNiSW N N203 3020 35.

\end{acknowledgements}

\bibliographystyle{aa}
\bibliography{14319}{}

\end{document}